# ADAPTIVE HYBRID SPEECH CODING WITH A MLP/LPC STRUCTURE

Marcos Faúndez-Zanuy

Escola Universitària Politècnica de Mataró.
Univertitat Politècnica de Catalunya
Avda. Puig i Cadafalch 101-111 08303 MATARO( BARCELONA, SPAIN)
e-mail:faundez@eupmt.es http://www.eupmt.es/veu
Tel:+34-93 757 44 04 Fax:+34-93-7570524

**ABSTRACT**
In the last years there has been a growing interest for nonlinear speech models. Several works have been published revealing the better performance of nonlinear techniques, but little attention has been dedicated to the implementation of the nonlinear model into real applications. This work is focused on the study of the behaviour of a combined linear/nonlinear predictive model based on linear predictive coding (LPC-10) and neural nets, in a speech waveform coder. Our novel scheme obtains an improvement in SEGSNR between 1 and 2.5 dB for an adaptive quantization ranging from 2 to 5 bits.

## 1. Introduction

Speech applications usually require the computation of a linear prediction model for the vocal tract. This model has been successfully applied during the last thirty years, but it has some drawbacks. Mainly, it is unable to model the nonlinearities involved in the speech production mechanism, and only one parameter can be fixed: the analysis order. With nonlinear models, the speech signal is better fit, and there is more flexibility to adapt the model to the application.

In the last years there has been a growing interest for nonlinear models applied to speech. This interest is based on the evidence of nonlinearities in the speech production mechanism. Several arguments justify this fact:
a) Residual signal of predictive analysis [1].
b) Correlation dimension of speech signal [2].
c) Fisiology of the speech production mechanism [3].
d) Probability density functions [4].
e) High order statistics [5].

Although these evidences, few applications have been developed so far. Mainly due to the high computational complexity and difficulty of analyzing the nonlinear systems.

The applications of the nonlinear predictive analysis have been focussed on speech coding, because it achieves greater prediction gains than LPC. The most relevant systems are [6] and [7], that have proposed a CELP with different nonlinear predictors that improve the SEGSNR of the decoded signal.

Three main approaches have been proposed for the nonlinear predictive analisys of speech. They are:

_______________________________________________________________________

a) Nonparametric prediction: it does not asume any model for the nonlinearity. It is a quite simple method, but the improvement over linear predictive methods is lower than with nonlinear parametric models.
b) Parametric prediction: it asumes a model of prediction. The main approaches are Volterra series and neural nets.
Recently several contributions have appeared on the context of neural nets. In this paper we propose a novel ADPCM speech waveform coder for the following bit rates: 16Kbps, 24Kbps, 32Kbps and 40Kbps with an hybrid (linear/nonlinear) predictor. With this structure a significative improvement in SEGSNR between 1 and 2.5 dB is achieved over the equivalent coders based on MLP and LPC alone.

## 2. Adaptive ADPCM with hybrid predictor scheme

A significative number of proposals found in the literature use Volterra series with quadratic nonlinearity (higher nonlinear functions imply a high number of coefficients and high computational burden for estimating them), and Radial Basis Function nets (RBF) that also implies a quadratic nonlinear model. We propose the use of a Multi Layer Perceptron net, because it has more flexibility in the nonlinearity. It is easy to show that an MLP with a sigmoid transfer function lets to model cubic nonlinearities (Taylor series expansion of sigmoid function). We believe that this is an important fact, because the nonlieraity present in the human speech prediction mechanism is due to a saturation

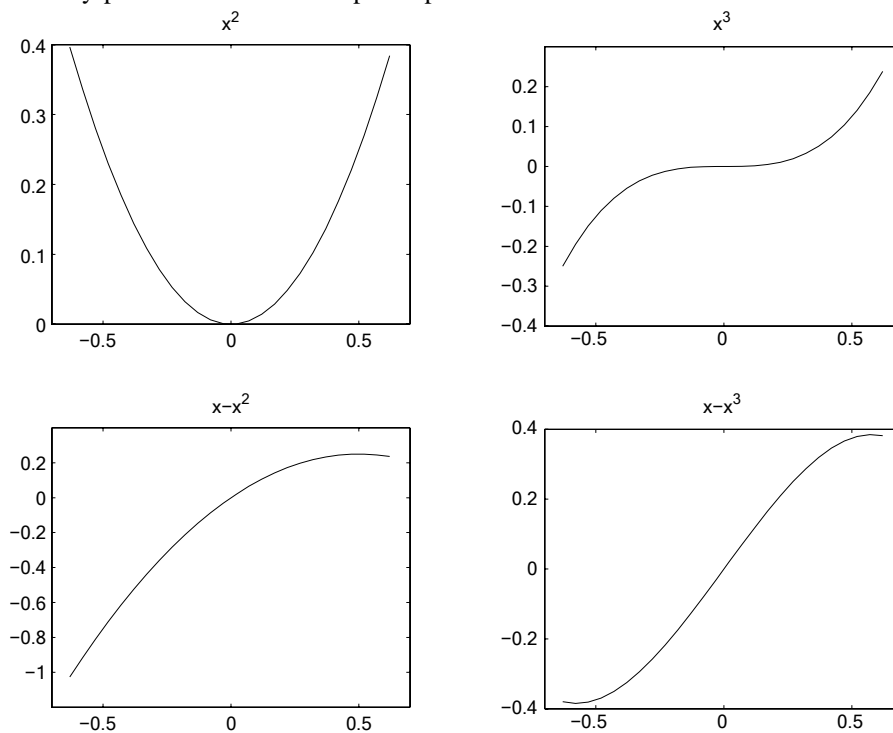

Figure 1 quadratic and cubic nonlinearity and the saturation function.

_______________________________________________________________________________

phenomena in the vocal chords. Figure 1 shows that is possible to model a saturation function with a cubic function, but it is not possible with a quadratic function.

A more detailed explanation about the nonlinear predictive model based on neural nets can be found in [8] and [9]. This paper is focused on the speech coding application.
In a preliminar work we studied the behaviour of the linear (LPC) and nonlinear MultiLayer Perceptron (MLP) predictors alone. This study reveals that the optimal solution is an adaptive selection LPC/MLP prediction. We propose a linear/non linear switched predictor in order to choose always the best predictor and to increase the SEGSNR of the decoded signal. Figure 2 represents the implemented scheme.
For each frame the outputs of the linear and nonlinear predictor are computed simultaneously with the coefficients obtained from the previous encoded frame. Then a logical decision is made that chooses the output with smaller prediction error. This implies an overhead of 1 bit for each frame that represents only 1/100 bits more per sample (in our simulations frame size is 100 samples). It is referred in the table as hybrid predictor, because it combines linear and nonlinear technologies. The percentage of use of each predictor is showed in table 1.

| PREDICTOR | Nq=2 | Nq=3 | Nq=4 | Nq=5 |
|---|---|---|---|---|
| LPC-10 | 60.54% | 54.07% | 54.13% | 52.75% |
| MLP | 39.46% | 45.93% | 45.87% | 47.25% |

table 1. Percentages of use LPC-10/MLP in the adaptive ADPCM backward speech coder

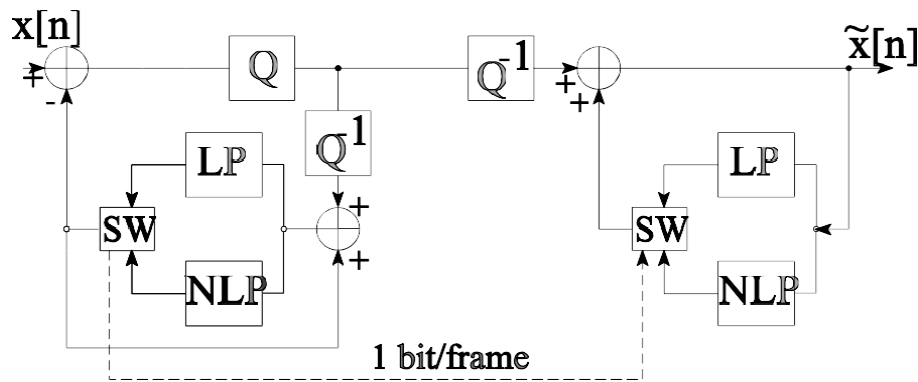

Fig. 2 Adaptive ADPCM-B hybrid coder. LP: linear predictor, NLP: nonlinear predictor, SW: switch

### 2.1 System overview

*Predictor coefficients updating*
 ! The coefficients are updated once time every frame.
 ! To avoid the transmission of the predictor coefficients an ADPCM backward

(ADPCMB) configuration is adopted. That is, the coefficients of the predictor are computed over the decoded previous frame, because it is already available at the receiver and it can compute the same coefficients values without any additional information. The obtained results with a forward unquantized predictor coefficients (ADPCMF) are also provided for comparison purposes.

! The nonlinear analysis consists on a multilayer perceptron with 10 input neurons, 2 hidden neurons and 1 output neuron. The network is trained with the Levenberg-Marquardt algorithm.

! The linear prediction analysis of each frame consists on 10 coefficients obtained with the autocorrelation method (LPC-10).

*Residual prediction error quantization*

! The prediction error has been quantized with (Nq=) 2 to 5 bits. (bit rate 16Kbps to 40Kbps).

! The quantizer step is adapted with multiplier factors, obtained from [10]. $\Delta_{max}$ and $\Delta_{min}$ are set empirically [11].

*Database*

! The results have been obtained with the following database: 8 speakers (4 males & 4 females) sampled at 8Khz and quantized at 12 bits/sample.

Additional details about the predictor and the database were reported in [8] and [9].

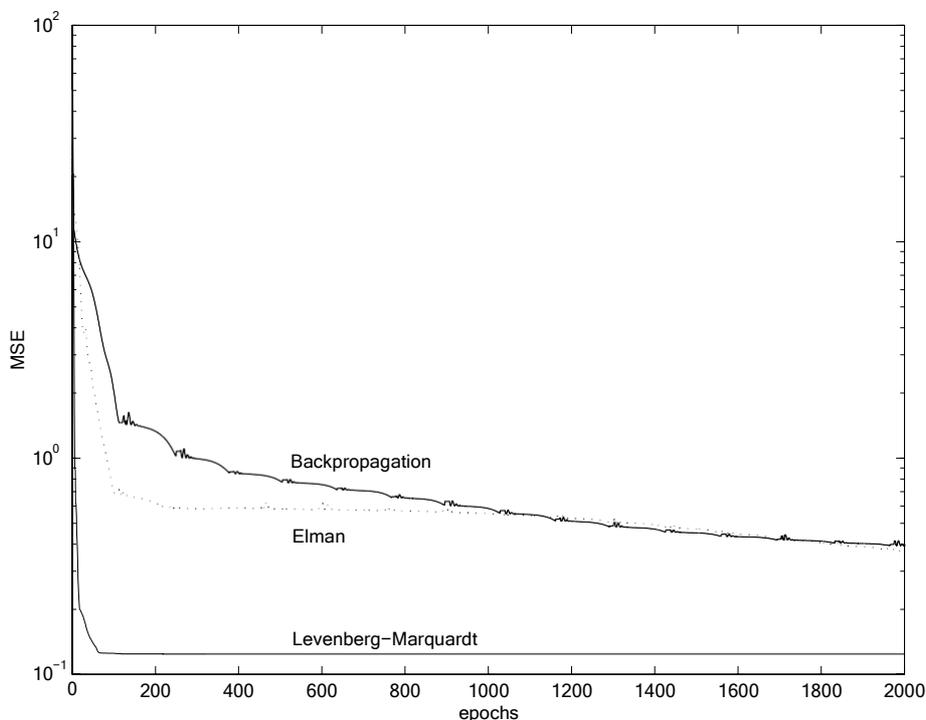

Figure 3 MSE vs epochs for Multilayer perceptron (trained with BP and L-M) and Elman net

### 2.2 Parameter selection

a) Linear predictor
For the linear predictor the parameters are:

! Prediction order: it is studied LPC-10 (same number of input samples than the MLP 10x2x1) and LPC-25 (same number of prediction coefficients than the MLP 10x2x1)
! Frame length: sizes from 10 to 300 samples with a step of 10 samples are evaluated. Notice that the bigger frame size the smaller the number of frames for a given speech signal, but if the frame length is large then the assumption of stationary signal into the analysis window is no valid and the behaviour degrades. If the frame length is short, the parameter estimation is not robust enough and the behaviour degrades.
b) Nonlinear predictor
For the nonlinear predictor based on neural nets, the number of parameters that must be optimized is greater. The selected network architecture is the Multi-Layer Perceptron with 10 input neurons, 2 hidden neurons with a sigmoid transfer function and one output neuron with a linear transfer function trained with the Levenberg-Marquardt (L-M) algorithm, based on our previous results [8]. We have also evaluated a recurrent Elman net, but we found that its behaviour was worse than MLP trained with L-M. Fig. 3 shows

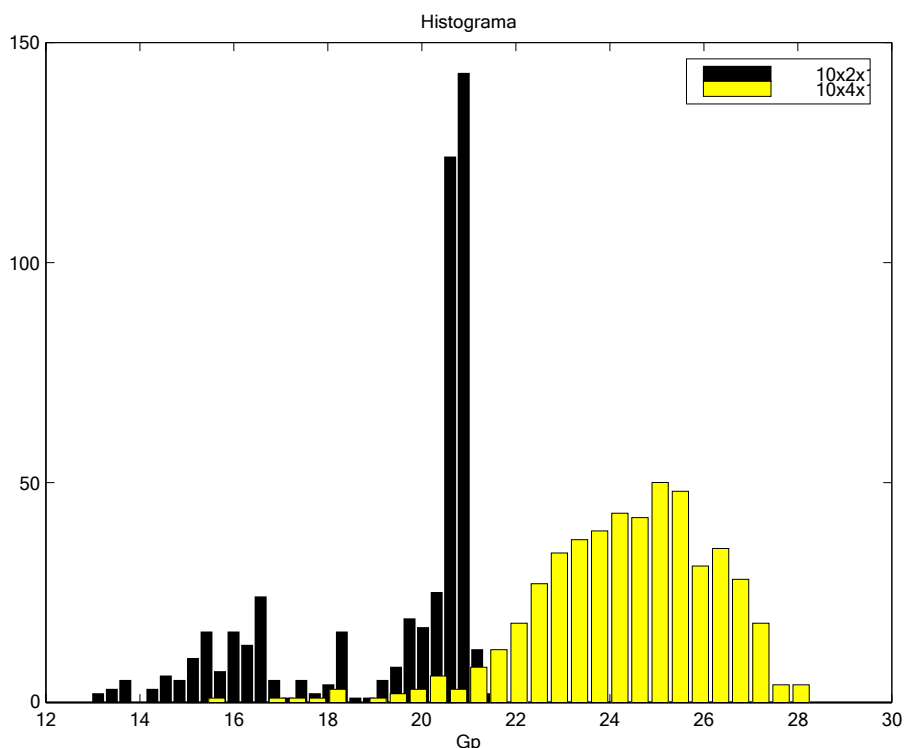

Figura 4 Histograms of the prediction gain for 500 random initializations of neural net weights.

the Mean Square Error as function of the number of epochs for a typical voiced frame of the database. It can be seen that the L-M algorithm presents a fast convergence and a small MSE. The MLP 10x4x1 was also tested, but it has more coefficients and the computational complexity is greater. Also a great number of random initializations must be done in the 10x4x1 structure, because the probability of achieving the greatest prediction gain for a random initialization is lower than for the 10x2x1 structure (fig. 4). The adjusted parameters of the predictor into the closed loop ADPCM scheme are:

! Number of trained epochs: This is a critical parameter. To encode a given frame the neural net is trained over the previous frame in the backward scheme and over the actual frame in the forward configuration. In both cases special attention must be taken in order to avoid the problem of overtraining (the network must have a good generalization capability to manage inputs not used for training). Although consecutive frames are normally very similar, there are significative changes in the waveform that must be seen as perturbances of the input, and even if the neural net is applied over the same frame used for training, the conditions are different because the predictor is trained in an open-loop scheme and tested in closed loop, so really the input signal is corrupted by the quantization noise. This is as much important the lesser is the number of quantizer bits. The way to make the neural net as robust as possible to this small changes implies the optimization of training conditions such us:

a) Number of epochs used for training
b) Number of random initializations of the weights ( a multistart algorithm is used).

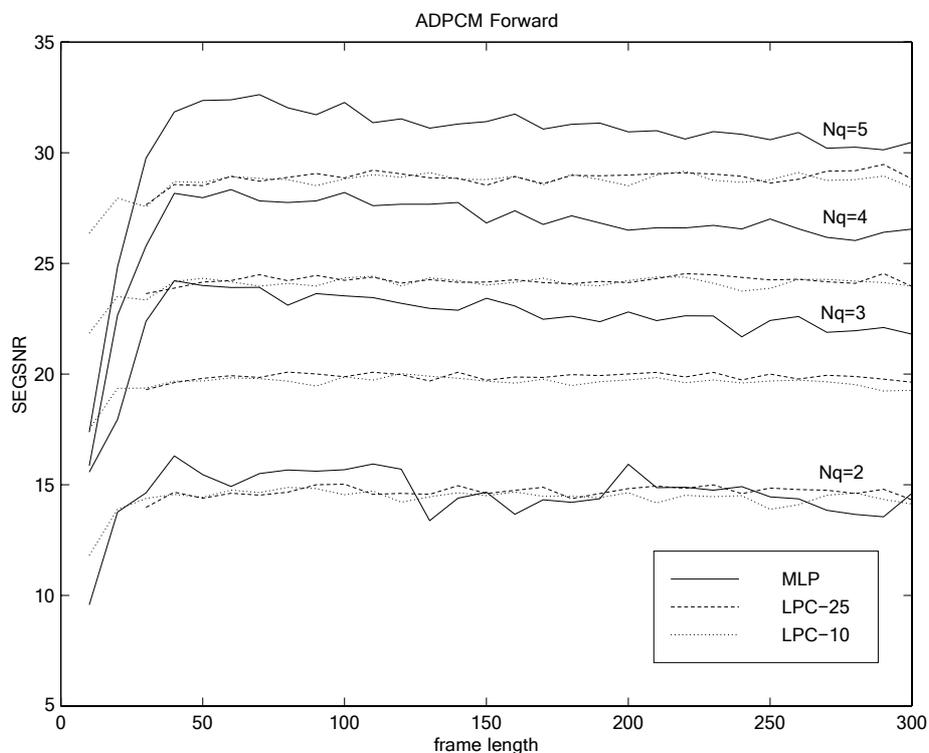

Fig. 5 SEGSNR vs frame length for ADPCM forward.

___________________________________________________________________

For achieving a good initialization a multi-start algorithm is used, which consists in computing several random initializations (experimentally fixed to 5) and to choose the one that achieves the higher SEGSNR. For selecting the number of epochs the optimal condition would be to evaluate for each frame the number of epochs that maximizes the SEGSNR. This is impractical because the decoder needs to know the number of epochs in order to track the encoder. Obviously this would imply the transmission of the number of trained epochs and so, the bit rate would be increased. The adopted solution consisted on a statistical study for choosing the best average number of epochs. This study reveals that the optimal number of epochs is 6 (see [9]).

! Frame length: Same commentaries of the linear predictor apply here. Experimental results show that the linear predictor has a similar behaviour over a wider range of frame sizes than the nonlinear predictor, but there is some rage for which the nonlinear predictor

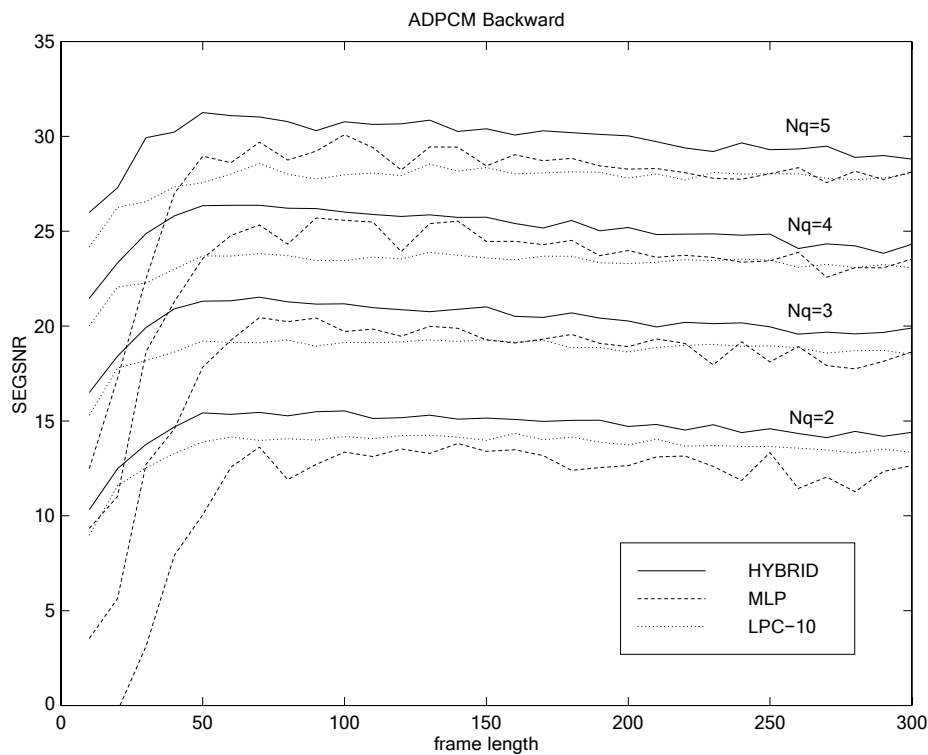

Fig. 6 SEGSNR vs frame length for ADPCM Backward
is better than the linear predictor.

Figures 5 and 6 show the SEGSNR (computed with a 200 samples analysis window) for frame lengths ranging from 10 to 300 samples for MLP10x2x1, LPC-10, LPC-25 and hybrid predictor with $N_q$=2 to 5 bits, averaged for the frames of one sentence. For the hybrid predictor an overhead of 1 bit/frame must be sent, so if the frame length is reduced the compression ratio is also reduced. For these reasons in this study the block size has

___

been selected to 100 samples/frame, because it offers a good compromise.

## 3. Results

The results have been evaluated using subjective criteria (listening to the original and decoded files), and SEGSNR.
Table 2 shows the SEGSNR obtained with the ADPCM configuration for the whole database with the following predictors: LPC-10, LPC-25 and MLP 10x2x1.
The results of the ADPCM forward (with unquantized predictor coefficients) are also provided such us reference of the backward configuration.
This results reveal the superiority of the nonlinear predictor in the forward configuration (3.5 dB aprox. over LPC-25 except for the 2 bit quantizer). This superiority is greater if the quantizer has a high number of levels.
In the backward configuration there is a small SEGSNR decrease with the linear predictor versus the forward configuration. For the nonlinear predictor it is more significative (nearly 3dB), but the SEGSNR is better than LPC-10 except for Nq=2 bits. Also, the variance of the SEGSNR is greater than for the linear predictor, because in the stationary portions of speech the neural net works satisfactorily well, and for the unvoiced parts the nnet generalizes poorly. Therefore, we propose a hybrid predictor.

| METHOD | Nq=2 bits | | Nq=3 bits | | Nq=4 bits | | Nq=5 bits | |
|---|---|---|---|---|---|---|---|---|
| | SEGSNR | std | SEGSNR | std | SEGSNR | std | SEGSNR | std |
| ADPCMF-LPC-10 | 15.35 | 5.8 | 21.18 | 6.4 | 25.86 | 6.9 | 30.52 | 7.1 |
| ADPCMF-LPC-25 | 15.65 | 5.6 | 21.46 | 6.4 | 26.26 | 6.9 | 30.79 | 7.2 |
| ADPCMF-MLP | 15.5 | 7.4 | 24.12 | 7.3 | 29.35 | 7.6 | 34.14 | 8.4 |
| ADPCMB-LPC-10 | 14.92 | 5.1 | 20.59 | 5.9 | 25.38 | 6.6 | 30.02 | 7.1 |
| ADPCMB-LPC-25 | 14.88 | 5.1 | 20.95 | 5.5 | 25.2 | 6 | 30.1 | 6.2 |
| ADPCMB-MLP | 14.35 | 6.9 | 21.48 | 7.5 | 26.76 | 7.6 | 31.5 | 8.4 |
| ADPCMB-HYBRID | 16.1 | 4.8 | 22.38 | 5.8 | 27.51 | 6.1 | 32.53 | 6.4 |

Table 2. SEGSNR for ADPCM forward, backward, linear, nonlinear and hybrid.

We have also evaluated the computational complexity of the studied systems. Table 3 summarizes the number of flops required for encoding the whole database with diferent schemes. For comparison purposes, the computational complexity has been refered to the ADPCM LPC-10 systems. Thus, the numbers in table 3 show how many times is greater the computational burden. Evaluated systems are:
- o  B: ADPCM with backward adaptation of prediction coefficients.

- o   F: ADPCM with forward adaptatition of unquantized prediction coefficients.
- o   L-10: linear predictive analisys of same order than MLP 10x2x1
- o   L-25: linear predictive analysis of same number of coefficients than MLP 10x2x1
- o   MLP: non linear predictive analisys with Multi Layer Perceptron 10x2x1
- o   H: Hybrid prediction (the best predictor, MLP 10x2x1 or LPC-10)

100 and 200 indicate the frame length in the block adaptive prediction system.

| scheme<br>frame length | BL10 | BL25 | BMLP | H | FL25 | FMLP |
|---|---|---|---|---|---|---|
| 100 | 1 | 1.4 | 27 | 29.8 | 1.4 | 24 |
| 200 | 2.1 | 2.6 | 27.5 | 32.9 | 2.6 | 26.2 |

Table 3: Computational burden

## 4. Conclusions and comparison with previously published work

The unique work that we have found that deals with ADPCM with nonlinear prediction is the one proposed by Mumolo et alt. [12]. It was based on Volterra series and has problems of unstability, which were overcome with a switched linear/nonlinear predictor. Our novel nonlinear scheme has been always stable in our experiments, although we also propose a switched predictor in order to increase the SEGSNR of the decoded signals. The results of our novel scheme show an increase between 1 and 2.5 dB over classical LPC-10 for quantizer ranges from 2 to 5 bits, while the work of Mumolo [12] is 1 dB over classical LPC for quantizer ranges from 3 to 4 bits and also with and hybrid predictor. On the other hand, the computational complexity has increased thirty times aproximately in the hybrid structure.

A statistical test was done in order to check if the results are statistically significatives. The selected test is ANOVA (Analysis of Variance), and it proves that the proposed adaptive hybrid speech coder is significatively better than the ADPCMB LPC-10 and LPC-25 schemes for all studied bit rates.

In this paper we have obatined the same conclusion than in our speaker recognition application of nonlinear predictive models based on MLP [13]: the best results are achieved with a combination of linear and nonlinear predictive models. In [14] we have obtained the same conclusion (also in speaker recognition) for a combination of a MLP trained as a classifier for each speaker, and a codebook of cepstral parameters derived from a linear parametrization.

## Acknowledgements

This work has been supported by the CICYT TIC97-1001-C02-02.

___